\documentclass[mathleft,letterpaper
]{an}
\usepackage{graphicx}
\usepackage{times}
\usepackage{natbib}
\bibpunct{(}{)}{;}{a}{}{,}
\begin{document}

\Pagespan{789}{}
\Yearpublication{2006}%
\Yearsubmission{2005}%
\Month{11}%
\Volume{999}%
\Issue{88}%

\newcommand{\Hii}{\ion{H}{ii} }
\newcommand{\Log}{\mbox{Log}}
\newcommand{\XMM}{XMM-{\em Newton}}
\newcommand{\xmm}{XMM-{\em Newton}}
\newcommand{\chandra}{{\em Chandra}}
\newcommand{\spitzer}{{\em Spitzer}}
\newcommand{\suzaku}{{\em Suzaku}}
\newcommand{\athena}{{\em Athena}}
\newcommand{\flux}{{\sc Flux} }
\newcommand{\lum}{{\sc Lum.} }
\newcommand{\lognlogs}{Log~$N$--Log~$S$}
\newcommand{\FIR}{{\rm FIR} }
\newcommand{\SFR}{{\rm SFR} }
\newcommand{\LX}{L_{\rm X} }
\newcommand{\Lsx}{L_{0.5-2} }
\newcommand{\Lhx}{L_{2-10} }
\newcommand{\de}{{\rm d}}
\newcommand{\ergs}{erg s$^{-1}$}
\newcommand{\ergscmq}{erg s$^{-1}$ cm$^{-2}$}
\newcommand{\ergscmqdegq}{erg s$^{-1}$ cm$^{-2}$ deg$^{-2}$}
\newcommand{\ergsHz}{erg s$^{-1}$ Hz$^{-1}$}
\newcommand{\phscmq}{photons s$^{-1}$ cm$^{-2}$}
\newcommand{\nWmqsr}{nW m$^{-2}$ sr$^{-1}$}
\newcommand{\e}[1]{\cdot 10^{#1}}
\newcommand{\figuraacolori}{{\em (In colour only in the electronic edition)}\/\ }
\newcommand{\fxott}{\Log (F_{{\rm X-ray}}/F_{{\rm opt}})}
\newcommand{\chiq}{\chi^2}
\newcommand{\chiqr}{\chi^2_{\rm red}}
\newcommand{\ael}{$\alpha$-elements}

\newcommand{\lesssim}{\la}   
\newcommand{\gtrsim}{\ga}

\newcommand{\cSvmekl}{{\tt c6vmekl}}
\newcommand{\cSpvmkl}{{\tt c6pvmkl}}
\newcommand{\cSpvmkT}{{\tt c6pvmk3}}
\newcommand{\tbabs}{{\tt tbabs}}
\newcommand{\ztbabs}{{\tt ztbabs}}
\newcommand{\pow}{{\tt pow}}

\title{Charge-exchange emission in the starburst galaxies M82 and
  NGC3256\thanks{Based on observations obtained with XMM-Newton, an ESA science mission with instruments and contributions directly funded by ESA Member States and NASA.}}

\author{P. Ranalli\inst{1,2}\fnmsep\thanks{Corresponding author:
  \email{piero.ranalli@oabo.inaf.it}\newline}
}
\titlerunning{Charge-exchange emission in M82 and NGC 3256}
\authorrunning{P. Ranalli}
\institute{
Universit\`a di Bologna, Dipartimento di Astronomia,
via Ranzani 1, 40127 Bologna, Italy
\and 
INAF--Osservatorio Astronomico di Bologna,
via Ranzani 1, 40127 Bologna, Italy
}

\received{14 Feb 2012}
\accepted{28 Feb 2012}
\publonline{later}


\abstract{%
  Charge-exchange (CE) emission produces features which are detectable
  with the current X-ray instrumentation in the brightest near
  galaxies.  We describe these aspects in the observed X-ray spectra
  of the star forming galaxies M82 and NGC 3256, from the \suzaku\ and
  \xmm\ telescopes. Emission from both ions (O, C) and neutrals (Mg,
  Si) is recognised. We also describe how microcalorimeter
  instrumentation on future missions will improve CE observations. }

\maketitle

\section{Introduction}
\label{sec:intro}

\sloppy

Charge-exchange emission (CE) has long been known to occur in optical
nebular spectra \citep{chamberlain56}. In the X-ray domain, it was
proposed for the first time to account for emission from the
comet Hyakutake \citep{cravens97}. Thereafter, it was invoked to
explain X-ray features from the solar system planets \citep[e.g., Mars
and Jupiter:][]{dennerl06,branduardi07}.  It has been suggested that it
might play a non-negligible role in explaining the emission from
galactic winds interacting with dense clouds \citep{lallement04}.

In the CE framework, ions from the wind diffuse into the cold gas; at
the interface, electrons are transferred from the neutrals to the
ions. The resulting ions can be highly excited, and
re-arrange their electrons by emitting photons in the extreme
ultraviolet and X-ray domains.  This process results only in line
emission and not in continuum. The photon emission rate is
proportional to the wind ion flux.

In external galaxies, CE might occur at the interface between the hot
wind \citep{heckman90} and the clouds of cold neutral gas within the
galaxy itself. Most of the emission is expected to come from ions of
the most abundant and light species present in galactic winds (O, C,
N; see R. Smith, this volume). It is also expected, if dust grains are
present and participate with the neutrals on their surface to the CE
process, that emission from the dust neutrals should be present;
considering olivine and pyroxene grains in the cold clouds, this would
resunt in lines from neutral Mg and Si \citep{djuric05}. Also, broad
H$\alpha$ emission could be detectable due to the CE providing thermal
balancing between protons in the superwind and neutral H atoms
(J. Raymond, this volume).

In this paper, we describe CE features in X-ray observations of the
star forming galaxies M82 (Sect.~\ref{sec:m82}) and NGC 3256
(Sect.~\ref{sec:n3256}). Finally, in Sect.~\ref{sec:future} we
prospect the advances from observations with near-future instruments.

\section{Charge-exchange in M82}
\label{sec:m82}

The nearby galaxy \object{M82} is often considered as the prototype starburst
in the local universe.  \citep{rie80}. At a distance of 3.63 Mpc
\citep{distanza_m82}, it has a luminosity of $10^{44}$ \ergs\ in the
far infrared and $10^{40}$ \ergs\ in the X-ray domain, from which a
Star Formation Rate (SFR) of $\sim 3$ M$_\odot$ yr$^{-1}$ can be
estimated.  The starburst is mainly located in the central regions of
the galaxy, and it is the origin of a large outflow perpendicular to
the plane of the galaxy and several kpc long \citep{heckman90}. The
outflow is visible at multiple wavelengths, such as the radio
\citep{seaquist91}, infrared \citep{alton99,engelbracht07,kaneda10},
H$\alpha$ and X-rays \citep{lehnert99}.  The observations in the X-ray
domain with the \chandra, \xmm\ and \suzaku\ observatories are
reported in \citet{grif00}, \citet{rs02}, \citet{or04},
\citet{strickland07}, \citet{tsuru07} and \citet{m82centok}. Thanks to
the small inclination of the galaxy disc, M82 lies almost edge-on,
thus allowing a good perspective on the outflow.

The CE was observed in M82 by both \suzaku\ \citep{tsuru07} and \xmm\
\citep{liu11,m82centok}.

\subsection{\suzaku\ observation}
\label{sec:suzaku}

\citet{tsuru07} found a marginal detection of the \ion{C}{VI} line at
0.459 keV with the XIS detector in the ``Cap'' region of M82. The Cap
is an area of extended X-ray emission, lying 11.6 kpc to the north of
the centre of M82, in the direction of the outflow. \citet{lehnert99}
suggested that it be the result of shock heating associated with an
encounter between the starburst-driven galactic superwind and a large
photoionized cloud in the halo of M82. The detected line, resulting
from a $n=4\rightarrow1$ transition, was attributed to CE interaction
of C ions with neutrals.

An upper limit to the contribution from CE to O line emission could be
drawn by considering the density of the H~I cloud in the Cap and the
cross-section of CE between an O ion and an H atom. Under the
hypothesis that all O ions undergo the CE with \ion{H}{I} and emit O~K
lines, the CE photon flux is estimated as $\lesssim 5\e{-6}$ \phscmq,
to be compared with an observed flux of $6\e{-6}$ \phscmq. This allows
the CE to be a contributor to the O lines. By the same logic, the
observed flux of \ion{C}{VI} is also consistent with a CE origin.

Altough they are not discussed in \citet{tsuru07}, lines from neutral
Mg and Si (see Sect.~\ref{sec:xmm}) were also detected in the spectrum
(T. Tsuru, priv.\ comm.), and indeed they are visible as residuals when
the Cap spectrum is fitted with a thermal model: either a collisional
ionization equilibrium one or a non-equilibrium one (Figs.~4 and 6 in
\citealt{tsuru07}).

\subsection{\xmm\ observations}
\label{sec:xmm}

\citet{m82centok} analyzed the X-ray spectra of several regions of the
M82 outflow, as part of a project to compare X-ray and stellar-based
chemical abundances \citep{or04}. The observation exposure was 100 ks
(73 ks after lightcurve cleaning); here we only discuss data from the
Reflection Grating Spectrometer (RGS). The RGS line spread function
(LSF) depends on the source shape and on the energy; because of
absorption, M82 has a different shape at the shorter wavelengths
(6--18 \AA) with respect to the longer ones ($\sim 22$ \AA). Thus it
is not possible to fit all the spectrum at once, but it must be
divided in chunks according to the LSF shape \citep[see][for
details]{m82centok}. Also, our analysis was done on all the RGS data,
which cover a few arcmin around the M82 centre. A study of the spatial
variation of the O line intensities was attempted by \citet[][see also
this volume]{liu11}, to whose paper we refer for details.

\begin{figure*}
  \centering
  \includegraphics[width=.3\textwidth,height=.39\textwidth]{Ovii-triplettosingolo.ps}
  \includegraphics[width=\columnwidth]{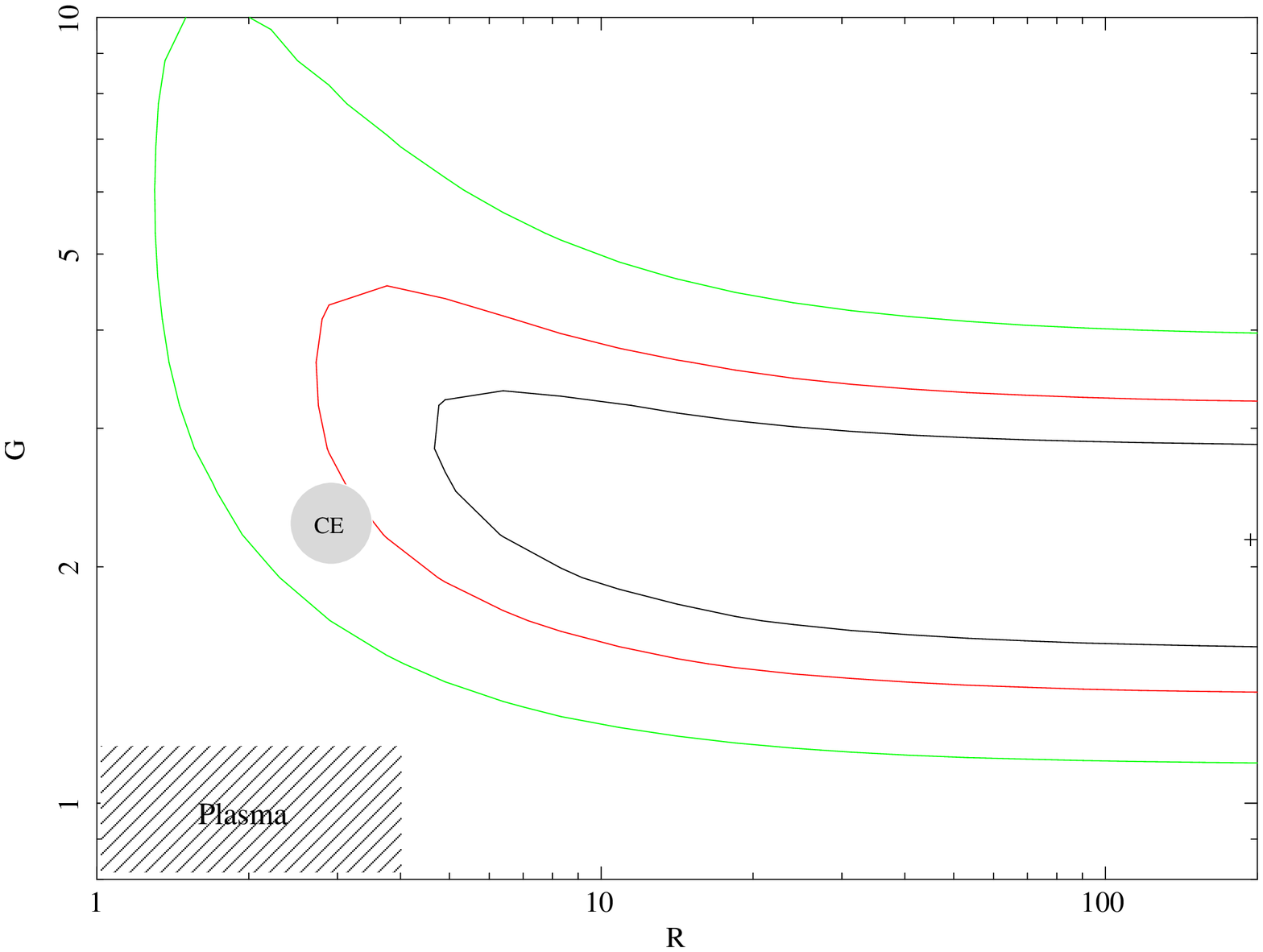}
  \caption{{\em Left panel:} RGS spectra and residuals of the O VII
    triplet with best-fitting model.  {\em
      Right panel:} Confidence contours (68.3\%, 90\% and 99\%) for
    the $R$ and $G$ parameters (which describe the line intensity
    ratios) for the O VII triplet. The spectrum is marginally
    consistent with CE emission (shown as the grey circle; the size of
    the circle is not representative of the uncertainties), but is not
    consistent with plasma emission (hashed rectangle). \textit{Both
      figures} are taken from \citet{m82centok}. }
  \label{fig:OVII}
\end{figure*}

\subsubsection*{Oxygen lines}
The \ion{O}{VII} triplet was observed at high spectral resolution
($E/\Delta E\sim 300$ at 1 keV for point sources ) with the RGS
spectrometer (Fig.~\ref{fig:OVII}, left panel). The triplet centroid
was found to be shifted towards the forbidden line, hinting for
CE-like line ratios. Altough the resolution is degraded by the large
extent of M82 and the lines are blended, the line ratios could still
be derived. By fitting the triplet with three narrow Gaussian lines,
the $R$ and $G$ parameters \citep{gabriel72} could be constrained to
$R\gtrsim 4$ and $1.5\lesssim G\lesssim 3$ (Fig.~\ref{fig:OVII}, right
panel). For comparison, laboratory measurements of CE line strength
resulted in $R\sim 3$ and $G\sim 2.2$ \citep{beiersdorfer03}; however,
the allowed values of $G$ could be much larger ($G \gtrsim 4$) because
of the smaller energy involved in the laboratory experiments with
respect to the what expected in astrophysical situations
(V. Krasnopolsky, priv.\ comm.). Conversely, the values expected for
ionization equilibrium, thermal emission are much different
($R\lesssim 4$, $G\lesssim 1.2$; \citealt{lallement04}) and at odds
with the RGS data. However, it is also possible that over-ionized
non-equilibrium plasma produce similar ratios. Although
\citet{m82centok} could not find an acceptable fit to the whole
spectrum with non-equilibrium plasma models, non-equilibrium might
still play a smaller role in a restricted part of the spetrum.
Overall, it is plausible that the processes of thermal and CE emission
contribute at similar levels.


Two \ion{O}{VIII} lines (at 16 \AA\ and 19 \AA) are strong enough to
be fitted individually in the RGS spectrum. The energies and profiles
of both lines can be modeled by thermal emission with the APEC model
\citep{apec}, after considering the appropriate line spread functions
(LSF). The MEKAL model \citep{mekal} leaves some residuals in the
15.8--16 \AA\ region, probably due to an incomplete modeling of the
\ion{Fe}{XVIII} lines at 15.4, 15.6 and 15.8 \AA.  By excluding one
line from the fit, O abundances can be derived from the other one, and
viceversa, thus allowing one to compare the abundances derived from one
single line. The abundances derived in this way are at odds with each
other: O/O$_\odot = 0.62\pm 0.10$ for the 16 \AA\ line, O/O$_\odot =
1.41\pm 0.05$ for the 19 \AA\ line (APEC model; in the Solar scale by
\citealt{grev98}).  The origin of this discrepancy is not clear,
basically because of potential systematic errors stemming from the
uncertainties in the modeling of the LSF, which for the RGS is both
energy- and source shape-dependent \citep[see][for
details]{m82centok}. If taken at their face values, the larger
abundance of the 19 \AA\ line might suggest the possibility of a
substantial contribution to its flux by CE. However, any verification
of this latter hypothesis, will likely have to wait for new telescopes
featuring microcalorimeter detectors (Astro-H, \athena), which will
not suffer by the LSF systematics that currently affect the RGS (see
Sect.~\ref{sec:future}).

\subsubsection*{Lines from neutral Mg and Si}

\citet{m82centok} report the detection of two lines at 7.2 and 10
\AA\ with the RGS instrument (with equivalent widths of 12 and 43 eV)
and of the 7.2 \AA\ line only with the EPIC instrument (equivalent
width in the 7--22 eV interval). The significance of the line
detections was assessed through Montecarlo simulations, which resulted
in the $>99.99\%$, 99.8\% and $>99.97\%$ confidence levels for the 7.2
and 10 \AA\ line in the RGS, and the 7.2 \AA\ line in the EPIC,
respectively.

These lines were identified with emission from neutral Mg at 9.92 \AA\
and neutral Si at 7.17 \AA. This interpretation is consistent with the
presence of olivine and pyroxene grains in the cold clouds
\citep{djuric05}, and is reinforced by the detection of infrared dust
emission in the M82 outflow \citep{kaneda10}.

\subsection{Discussion}
\label{sec:m82disc}

The evidence for CE phenomena in M82 relies on three observations:
i) non-thermal $R$ and $G$ line ratios in the \ion{O}{VII} RGS spectrum;
ii) lines from neutral Mg and Si in RGS and EPIC;
iii) marginal detection of C VI in the Cap spectrum with \suzaku.
The strongest claim for CE here is probably in the first one ---
though the evidence for CE is not yet univoque.  The other two claims
should be rather regarded as further evidence in the framework where
the \ion{O}{VII} line ratios have been attributed to CE. More work is
needed, both on the theoretical side (more reliable predictions of the
line ratios) and on the observational side (higher resolution data),
before these observations can be regarded as a \textit{proof} of CE.

\section{Charge-exchange in NGC 3256}
\label{sec:n3256}

\object{NGC 3256} is a luminous dusty merger remnant, with a SFR about
10 times larger than M82. In antithesis to M82, NGC 3256 lies face-on,
so that the galactic outflow points towards us and is superimposed on
the galactic disc.  The 3 and 6\,cm radio maps \citep{norrisforbes95}
reveal two distinct, resolved (FWHM $\sim1.2\arcsec$) nuclei and some
fainter diffuse radio emission. Separated by $5\arcsec$ in
declination, the two cores dominate the radio emission, the northern
one being slightly (15\%) brighter. The northern core is also the
brightest spot in X-rays, while the southern one lies behind a dust
lane and is only visible at energies $\gtrsim 2$ keV. Several
point-sources and bright diffuse emission are also present
\citep{rcs03,lira}.

\begin{figure}
  \centering
  \includegraphics[height=\columnwidth,angle=-90]{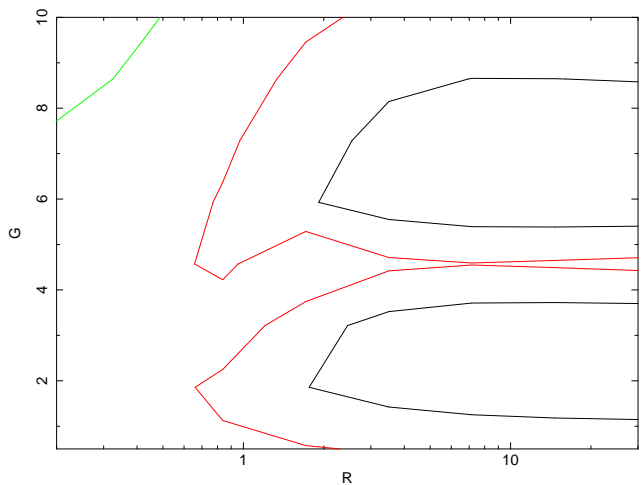}
  \caption{ Confidence contours for the O VII triplet in NGC 3256. 
    Two different regions for $G$ seem to be allowed; both are
    consistent with CE expectations. This triplet is added \textit{on
      top of the thermal emission}, contrarily to what done in
    Fig.~\ref{fig:OVII}.
  }
  \label{fig:OVIIngc3256}
\end{figure}

\begin{sloppypar}
NGC 3256 was observed by \xmm\ for 130 ks, with 103 ks remaining after
cleaning for background flares. The spectrum was modeled as thermal
emission plus a power-law to account for the X-ray binaries. The LSF
for NGC 3256 is narrower and less energy-dependent than for M82,
because of the smaller source extent of the former. Thus it was
possible to fit the whole RGS spectrum together. We considered a
multi-temperature thermal model for the plasma, plus a power-law for
the X-ray binaries. This fitted well the observed spectrum, including
the \ion{O}{VIII} lines, but it underestimated the \ion{O}{VII}
triplet at $\sim 22$\AA. Thus, a set of three narrow Gaussian lines at
the \ion{O}{VII} energies was added. The resulting parameters for the
line intensity ratios are shown in Fig.~\ref{fig:OVIIngc3256}: they
show that a CE component is detected \textit{on top of the thermal
  emission}, notwithstanding the uncertainties on the $G$ parameter.
\end{sloppypar}

Some residuals are visible at the energies of the neutral Mg and Si
lines, though a proper analysis of their significance has not yet been
made. A paper is in preparation on this subject.

\section{Microcalorimeters in future missions}
\label{sec:future}

Significant advances in the study of CE emission in external galaxies
are expected with the future X-ray missions which will feature
microcalorimeter detections: both the JAXA-led Astro-H, with a launch date
set for year 2013, and \athena, currently under discussion at
ESA. This technology was briefly demonstrated with the \suzaku/XRS
instrument. On orbit, the XRS achieved a 7 eV resolution at 6 keV
\citep{kelley07} when observing the on-board calibration
source. However, its cryostat failed just two weeks after the
launch,  rendering the XRS unusable for astronomical observations.

Microcalorimeters provide a resolution, at energies $\lesssim 2$ keV
which are of interest for CE phenomena, of the same order of the
nominal one of the \xmm\ RGS for point sources, but without any
complication due to the slitless nature of the grating spectrometers
such as the RGS, or the \chandra\ LETGS and HETGS (note, if comparing
with Fig.~\ref{fig:OVII}, that for M82 the RGS resolution is severely
degraded by its large dimensions on the sky). Microcalorimeters also
provide spatial resolution, allowing one, for example, to remove detected
binaries.

\begin{figure}
  \centering
  \includegraphics[height=\columnwidth,angle=-90]{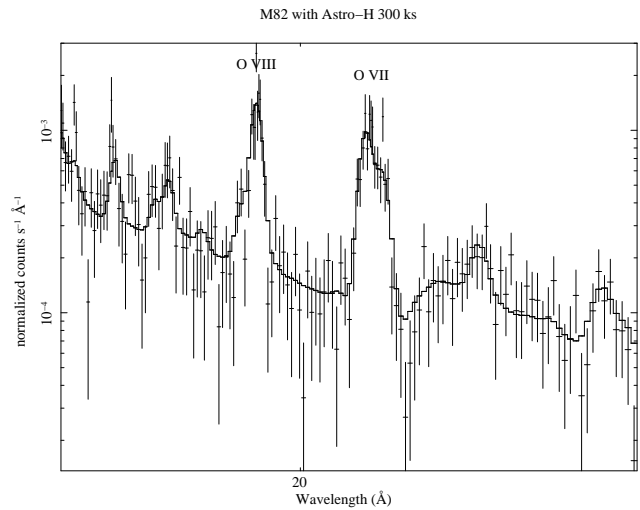}
  \caption{ Simulation of the M82 spectrum with Astro-H, showing the O
    lines at 19 and $\sim 22$ \AA. A thermal model is assumed for the
    calculation (visible from the \ion{O}{VII} line shape).
  }
  \label{fig:calorimeter}
\end{figure}

A simulation of the M82 spectrum as observed with the SXS
microcalorimeter which is part of the Astro-H instruments is shown in
Fig.~\ref{fig:calorimeter}. For simplicity, the model is thermal, with
no contribution from CE; this is noticeable from the \ion{O}{VII} line
shape. The simulated exposure time is 300 ks. 

Astro-H, however, has a very large PSF ($\sim 1$ arcmin) and will not
be able to remove X-ray binaries. \athena, with a planned PSF of $\sim
5$ arcsec, should be able to remove at least the brightest ones, such
as M82-X1 \citep[see Fig.1 in][]{m82centok}, thus enhancing the
signal/background ratio for the plasma component of the spectrum.

\acknowledgements 
\begin{sloppypar}
We warmly thank the Institut d'Astro\-phy\-sique de Paris (IAP) for
support, and T. Montmerle for his kind hospitality.  Financial
contribution from the agreement ASI-INAF I/009/10/0 is acknowledged.
\end{sloppypar}

%
%
%

\def\aj{AJ}%
\def\araa{ARA\&A}%
\def\apj{ApJ}%
\def\apjl{ApJ}%
\def\apjs{ApJS}%
\def\ao{Appl.~Opt.}%
\def\apss{Ap\&SS}%
\def\aap{A\&A}%
\def\aapr{A\&A~Rev.}%
\def\aaps{A\&AS}%
\def\azh{AZh}%
\def\baas{BAAS}%
\def\jrasc{JRASC}%
\def\memras{MmRAS}%
\def\mnras{MNRAS}%
\def\pra{Phys.~Rev.~A}%
\def\prb{Phys.~Rev.~B}%
\def\prc{Phys.~Rev.~C}%
\def\prd{Phys.~Rev.~D}%
\def\pre{Phys.~Rev.~E}%
\def\prl{Phys.~Rev.~Lett.}%
\def\pasp{PASP}%
\def\pasj{PASJ}%
\def\qjras{QJRAS}%
\def\skytel{S\&T}%
\def\solphys{Sol.~Phys.}%
\def\sovast{Soviet~Ast.}%
\def\ssr{Space~Sci.~Rev.}%
\def\zap{ZAp}%
\def\nat{Nature}%
\def\iaucirc{IAU~Circ.}%
\def\aplett{Astrophys.~Lett.}%
\def\apspr{Astrophys.~Space~Phys.~Res.}%
\def\bain{Bull.~Astron.~Inst.~Netherlands}%
\def\fcp{Fund.~Cosmic~Phys.}%
\def\gca{Geochim.~Cosmochim.~Acta}%
\def\grl{Geophys.~Res.~Lett.}%
\def\jcp{J.~Chem.~Phys.}%
\def\jgr{J.~Geophys.~Res.}%
\def\jqsrt{J.~Quant.~Spec.~Radiat.~Transf.}%
\def\memsai{Mem.~Soc.~Astron.~Italiana}%
\def\nphysa{Nucl.~Phys.~A}%
\def\physrep{Phys.~Rep.}%
\def\physscr{Phys.~Scr}%
\def\planss{Planet.~Space~Sci.}%
\def\procspie{Proc.~SPIE}%
\let\astap=\aap
\let\apjlett=\apjl
\let\apjsupp=\apjs
\let\applopt=\ao
\bibliographystyle{aa}
\bibliography{../fullbiblio}

\begin{thebibliography}{30}
\expandafter\ifx\csname natexlab\endcsname\relax\def\natexlab#1{#1}\fi

\bibitem[{{Alton} {et~al.}(1999){Alton}, {Davies}, \& {Bianchi}}]{alton99}
{Alton}, P.~B., {Davies}, J.~I., \& {Bianchi}, S. 1999, \aap, 343, 51

\bibitem[{{Beiersdorfer} {et~al.}(2003){Beiersdorfer}, {Boyce}, {Brown},
  {Chen}, {Kahn}, {Kelley}, {May}, {Olson}, {Porter}, {Stahle}, \&
  {Tillotson}}]{beiersdorfer03}
{Beiersdorfer}, P., {Boyce}, K.~R., {Brown}, G.~V., {et~al.} 2003, Science,
  300, 1558

\bibitem[{{Branduardi-Raymont} {et~al.}(2007){Branduardi-Raymont}, {Bhardwaj},
  {Elsner}, {Gladstone}, {Ramsay}, {Rodriguez}, {Soria}, {Waite}, \&
  {Cravens}}]{branduardi07}
{Branduardi-Raymont}, G., {Bhardwaj}, A., {Elsner}, R.~F., {et~al.} 2007, \aap,
  463, 761

\bibitem[{{Chamberlain}(1956)}]{chamberlain56}
{Chamberlain}, J.~W. 1956, \apj, 124, 390

\bibitem[{{Cravens}(1997)}]{cravens97}
{Cravens}, T.~E. 1997, \grl, 24, 105

\bibitem[{{Dennerl} {et~al.}(2006){Dennerl}, {Lisse}, {Bhardwaj}, {Burwitz},
  {Englhauser}, {Gunell}, {Holmstr{\"o}m}, {Jansen}, {Kharchenko}, \&
  {Rodr{\'{\i}}guez-Pascual}}]{dennerl06}
{Dennerl}, K., {Lisse}, C.~M., {Bhardwaj}, A., {et~al.} 2006, \aap, 451, 709

\bibitem[{{Djuri{\'c}} {et~al.}(2005){Djuri{\'c}}, {Lozano}, {Smith}, \&
  {Chutjian}}]{djuric05}
{Djuri{\'c}}, N., {Lozano}, J.~A., {Smith}, S.~J., \& {Chutjian}, A. 2005,
  \apj, 635, 718

\bibitem[{{Engelbracht} {et~al.}(2006){Engelbracht}, {Kundurthy}, {Gordon},
  {Rieke}, {Kennicutt}, {Smith}, {Regan}, {Makovoz}, {Sosey}, {Draine},
  {Helou}, {Armus}, {Calzetti}, {Meyer}, {Bendo}, {Walter}, {Hollenbach},
  {Cannon}, {Murphy}, {Dale}, {Buckalew}, \& {Sheth}}]{engelbracht07}
{Engelbracht}, C.~W., {Kundurthy}, P., {Gordon}, K.~D., {et~al.} 2006, \apjl,
  642, L127

\bibitem[{{Freedman} {et~al.}(1994){Freedman}, {Hughes}, {Madore}, {Mould},
  {Lee}, {Stetson}, {Kennicutt}, {Turner}, {Ferrarese}, {Ford}, {Graham},
  {Hill}, {Hoessel}, {Huchra}, \& {Illingworth}}]{distanza_m82}
{Freedman}, W.~L., {Hughes}, S.~M., {Madore}, B.~F., {et~al.} 1994, \apj, 427,
  628

\bibitem[{{Gabriel} \& {Jordan}(1972)}]{gabriel72}
{Gabriel}, A.~H. \& {Jordan}, C. 1972, in {Case studies in atomic collision
  physics}, ed. {McDaniel} \& {McDowell} No.~2, 209

\bibitem[{{Grevesse} \& {Sauval}(1998)}]{grev98}
{Grevesse}, N. \& {Sauval}, A.~J. 1998, Space Science Reviews, 85, 161

\bibitem[{{Griffiths} {et~al.}(2000){Griffiths}, {Ptak}, {Feigelson},
  {Garmire}, {Townsley}, {Brandt}, {Sambruna}, \& {Bregman}}]{grif00}
{Griffiths}, R.~E., {Ptak}, A., {Feigelson}, E.~D., {et~al.} 2000, Science,
  290, 1325

\bibitem[{{Heckman} {et~al.}(1990){Heckman}, {Armus}, \& {Miley}}]{heckman90}
{Heckman}, T.~M., {Armus}, L., \& {Miley}, G.~K. 1990, \apjs, 74, 833

\bibitem[{{Kaneda} {et~al.}(2010){Kaneda}, {Ishihara}, {Suzuki}, {Ikeda},
  {Onaka}, {Yamagishi}, {Ohyama}, {Wada}, \& {Yasuda}}]{kaneda10}
{Kaneda}, H., {Ishihara}, D., {Suzuki}, T., {et~al.} 2010, \aap, 514, A14

\bibitem[{{Kelley} {et~al.}(2007){Kelley}, {Mitsuda}, {Allen}, {Arsenovic},
  {Audley}, {Bialas}, {Boyce}, {Boyle}, {Breon}, {Brown}, {Cottam}, {Dipirro},
  {Fujimoto}, {Furusho}, {Gendreau}, {Gochar}, {Gonzalez}, {Hirabayashi},
  {Holt}, {Inoue}, {Ishida}, {Ishisaki}, {Jones}, {Keski-Kuha}, {Kilbourne},
  {McCammon}, {Morita}, {Moseley}, {Mott}, {Narasaki}, {Ogawara}, {Ohashi},
  {Ota}, {Panek}, {Porter}, {Serlemitsos}, {Shirron}, {Sneiderman},
  {Szymkowiak}, {Takei}, {Tveekrem}, {Volz}, {Yamamoto}, \&
  {Yamasaki}}]{kelley07}
{Kelley}, R.~L., {Mitsuda}, K., {Allen}, C.~A., {et~al.} 2007, \pasj, 59, 77

\bibitem[{{Lallement}(2004)}]{lallement04}
{Lallement}, R. 2004, \aap, 422, 391

\bibitem[{{Lehnert} {et~al.}(1999){Lehnert}, {Heckman}, \&
  {Weaver}}]{lehnert99}
{Lehnert}, M.~D., {Heckman}, T.~M., \& {Weaver}, K.~A. 1999, \apj, 523, 575

\bibitem[{{Lira} {et~al.}(2002){Lira}, {Ward}, {Zezas}, {Alonso-Herrero}, \&
  {Ueno}}]{lira}
{Lira}, P., {Ward}, M., {Zezas}, A., {Alonso-Herrero}, A., \& {Ueno}, S. 2002,
  \mnras, 330, 259

\bibitem[{{Liu} {et~al.}(2011){Liu}, {Mao}, \& {Wang}}]{liu11}
{Liu}, J., {Mao}, S., \& {Wang}, Q.~D. 2011, \mnras, 415, L64

\bibitem[{{Mewe} {et~al.}(1995){Mewe}, {Kaastra}, \& {Liedahl}}]{mekal}
{Mewe}, R., {Kaastra}, J.~S., \& {Liedahl}, D.~A. 1995, \apj, 6

\bibitem[{{Norris} \& {Forbes}(1995)}]{norrisforbes95}
{Norris}, R.~P. \& {Forbes}, D.~A. 1995, \apj, 446, 594

\bibitem[{{Origlia} {et~al.}(2004){Origlia}, {Ranalli}, {Comastri}, \&
  {Maiolino}}]{or04}
{Origlia}, L., {Ranalli}, P., {Comastri}, A., \& {Maiolino}, R. 2004, \apj,
  606, 862

\bibitem[{{Ranalli} {et~al.}(2008){Ranalli}, {Comastri}, {Origlia}, \&
  {Maiolino}}]{m82centok}
{Ranalli}, P., {Comastri}, A., {Origlia}, L., \& {Maiolino}, R. 2008, \mnras,
  386, 1464

\bibitem[{{Ranalli} {et~al.}(2003){Ranalli}, {Comastri}, \& {Setti}}]{rcs03}
{Ranalli}, P., {Comastri}, A., \& {Setti}, G. 2003, \aap, 399, 39

\bibitem[{{Read} \& {Stevens}(2002)}]{rs02}
{Read}, A.~M. \& {Stevens}, I.~R. 2002, \mnras, 335, L36

\bibitem[{{Rieke} {et~al.}(1980){Rieke}, {Lebofsky}, {Thompson}, {Low}, \&
  {Tokunaga}}]{rie80}
{Rieke}, G.~H., {Lebofsky}, M.~J., {Thompson}, R.~I., {Low}, F.~J., \&
  {Tokunaga}, A.~T. 1980, \apj, 238, 24

\bibitem[{{Seaquist} \& {Odegard}(1991)}]{seaquist91}
{Seaquist}, E.~R. \& {Odegard}, N. 1991, \apj, 369, 320

\bibitem[{{Smith} {et~al.}(2001){Smith}, {Brickhouse}, {Liedahl}, \&
  {Raymond}}]{apec}
{Smith}, R.~K., {Brickhouse}, N.~S., {Liedahl}, D.~A., \& {Raymond}, J.~C.
  2001, \apjl, 556, L91

\bibitem[{{Strickland} \& {Heckman}(2007)}]{strickland07}
{Strickland}, D.~K. \& {Heckman}, T.~M. 2007, \apj, 658, 258

\bibitem[{{Tsuru} {et~al.}(2007){Tsuru}, {Ozawa}, {Hyodo}, {Matsumoto},
  {Koyama}, {Awaki}, {Fujimoto}, {Griffiths}, {Kilbourne}, {Matsushita},
  {Mitsuda}, {Ptak}, {Ranalli}, \& {Yamasaki}}]{tsuru07}
{Tsuru}, T.~G., {Ozawa}, M., {Hyodo}, Y., {et~al.} 2007, \pasj, 59, 269

\end{thebibliography}

\end{document}